\documentclass[preprint,noshowpacs,preprintnumbers,amsmath,amssymb]{revtex4}

\usepackage{graphicx}
\usepackage{dcolumn}
\usepackage{bm}
\usepackage{color}

\begin{document}

\title{Interlayer tunneling in counterflow experiments on the excitonic condensate in quantum Hall bilayers}

\author{Y. Yoon}
\author{L. Tiemann}
\altaffiliation[Now at ]{NTT Basic Research Laboratories, Japan}
\author{S. Schmult}
\author{W. Dietsche}
\author{K. von Klitzing}
\affiliation{ Max-Planck-Institut f\"{u}r Festk\"{o}rperforschung,
Heisenbergstra{\ss}e
1, 70569 Stuttgart, Germany}
\author{W. Wegscheider}
\altaffiliation[Now at ]{ETH Z\"{u}rich, Switzerland}
\affiliation{Institut f\"{u}r Experimentelle und Angewandte Physik, Universit\"{a}t Regensburg, D-93040 Regensburg, Germany}

\begin{abstract}
{The effect of tunneling on the transport properties of} quantum Hall double layers in the regime of the excitonic condensate at total filling factor one is studied in counterflow experiments. If
 the tunnel current $I$ is smaller than a critical $I_C$, tunneling is large and is effectively shorting the two layers. For $I > I_C$ tunneling becomes negligible. Surprisingly, the transition between the two tunneling regimes has only a minor impact on the features of the filling-factor one state as observed in magneto-transport, but at currents exceeding $I_C$ the resistance along the layers increases rapidly.
\end{abstract}

\maketitle

Interest in the Bose-Einstein condensation of excitons was revived with the advent of independently contacted two-layer quantum Hall systems with a small layer separation. Under a perpendicular magnetic field  $B$, the density of states in the two 2-dimensional  electron-gases (2DEGS) separates into Landau levels. The  filling factor $\nu$ ($\nu=hn/eB$, $n$: electron density) determines the occupation of these levels. If the filling factor in each layer is adjusted to 1/2 and if  the magnetic length $l_B={\sqrt{\hbar/eB}}$ becomes comparable to the layer separation, Coulomb interaction favors the emergence of a correlated bilayer state, generally known as the total filling factor one ($\nu_T$=1) state. Excitons are thought to form between the occupied and the unoccupied (hole) states in the neighboring layers, and there is strong evidence for their condensation into a macroscopic quantum state at very low temperatures. 
  
It had been proposed that this state has a  BCS-like ground state \cite {Fer1989, Mac1990, Eis2004} suggesting the existence of an excitonic BCS superfluid. Indeed several remarkable properties have been observed such as an enhanced interlayer (tunnel)  conductance \cite {Spi2000, Wie2006,Fin2008}. Recently, even a nearly dissipationless interlayer  tunneling current has been detected which terminates upon reaching a critical value \cite {Tie2008}. Such a Josephson-like behavior has also been theoretically predicted \cite {Wen1993,Ezawa1994,Par2006,Dolcini2010}. In counterflow experiments where the current flows through one layer and is then redirected in opposite direction through the other layer, a nearly complete vanishing of both the longitudinal ($R_{xx}$) and the Hall ($R_{xy}$) resistance has been observed \cite {Kell2004, Tut2004, Wie2004}. This has been taken as evidence for superfluidity-like features of the excitonic condensate. Interestingly, only a negligible ''leakage'' of the current was found in the reported counterflow experiments which appears to be in conflict with the just-mentioned observation of a nearly dissipationless Josephson-like current. 
 
In this Letter, we present new results in the counterflow set-up on double-layer electron systems which show that below a critical current strong tunneling effectively short-circuits both layers. Above the critical value,   the tunneling is strongly suppressed. Surprisingly, the transition between these two tunneling regimes has only a minor impact on the features of the $\nu_T$=1 state in magneto-transport causing only a small increase of the counterflow longitudinal resistance at $\nu_T$=1.  We find, however, that the dissipation of the longitudinal transport increases rapidly in the $\nu_T$=1 state after the critical current is exceeded.

The double quantum well system used in this study has been grown by molecular beam epitaxy. Two 19~nm GaAs wells are separated by a 9.6 nm superlattice barrier layer consisting of alternating layers of AlAs (1.7~nm) and GaAs (0.28~nm). The quantum wells are populated by Si-doped sheets placed 300 nm below and 280 nm above the wells, yielding intrinsic densities of approximately 4.0$\times$10$^{10}~$cm$^{-2}$ in the upper and 4.6$\times$10$^{10}~$cm$^{-2}$ in the lower quantum well. Mobilities at these densities are 4.3$\times$10$^{5}~$cm$^{2}$V$^{-1}$s$^{-1}$ and 4.1$\times$10$^{5}~$cm$^{2}$V$^{-1}$s$^{-1}$, respectively. A Hall bar device with a length of 880 $\mu$m and a width of 80 $\mu$m is prepared by optical lithography. Independent electrical contacts to the two layers are achieved by applying appropriate negative voltages to the pre-patterned buried back gates and the metallic front gates crossing the contact arms (selective depletion technique \cite {Eis1990, Rub1998}). With the intrinsic densities, the interlayer resistance at zero magnetic field and 4.2~K is around 380 k$\Omega$. The densities in the two layers can be adjusted independently by  additional front and back gates and are balanced in all data shown. The ratio ($d/l_B$) of the layer separation ($d=28.6~nm$) to the magnetic length $l_B$ is set to values between 1.37 and 1.68.  All data were taken in a dilution refrigerator at a bath temperature of 20 - 30 mK.

\begin{figure}
  \includegraphics {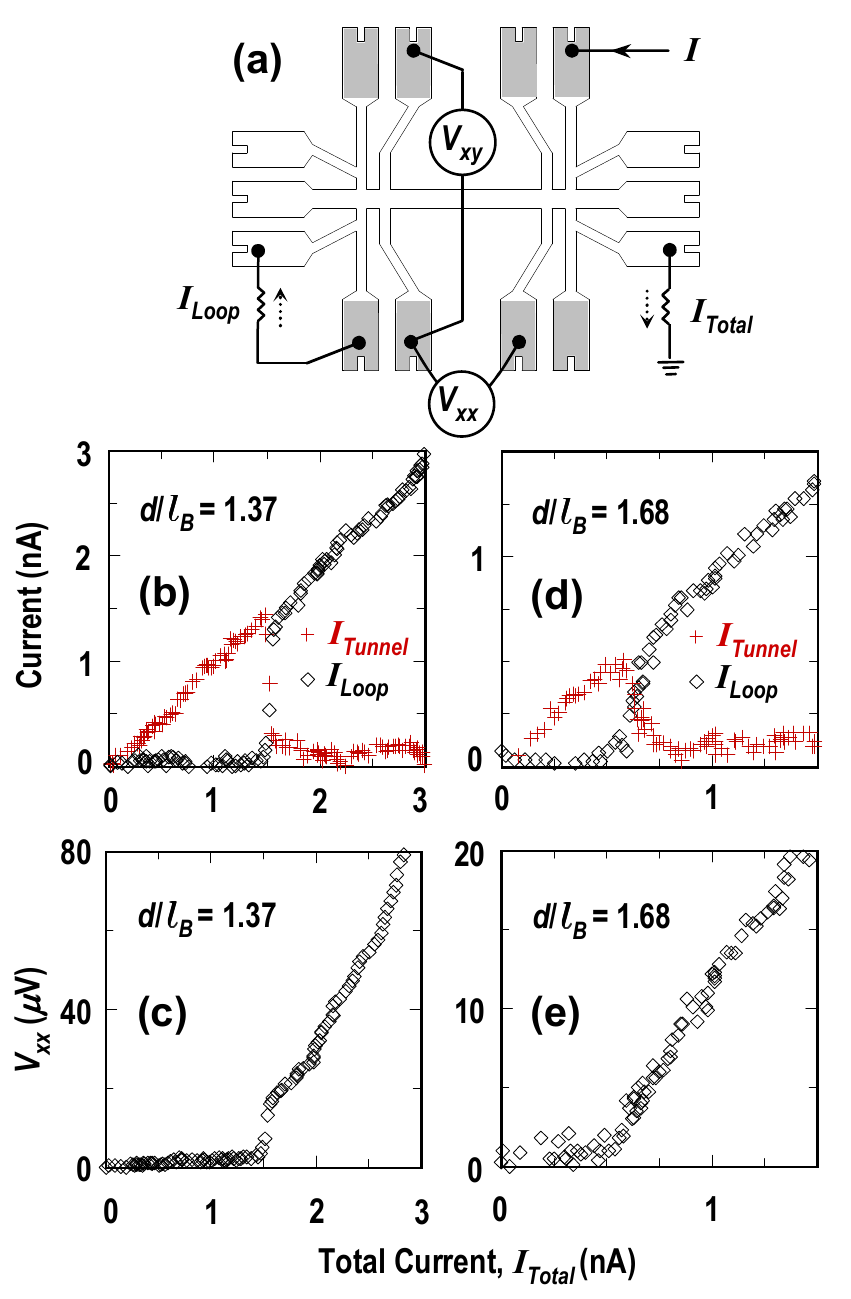}\\
  \textbf{\caption{\label{fig:1} \textbf{(a) A schematic of the measurement set-up. {The $\nu_T$ = 1 state forms along the elongated part of the structure by adjusting the densities using front and backgate (not shown).} Shaded contacts connect to the top layer. $I$ is first injected into the top layer, extracted and then redirected via a loop resistor (10 k$\Omega$) into the bottom layer, and finally guided via a total resistor (10 k$\Omega$) into ground. The currents $I_{Total}$ and $I_{Loop}$ are obtained from the voltage drops across the respective resistors. $I_{Tunnel}$ is defined as $I_{Total}$ - $I_{Loop}$.  At $d/l_B$=1.37 (b and c) and $d/l_B$=1.68 (d and e) $I_{Tunnel}$ \& $I_{Loop}$ and $V_{xx}$ are plotted versus $I_{Total}$.}}}
\end{figure}

Counterflow experiments are performed by passing two oppositely directed currents through the two layers as shown in Fig.~\ref{fig:1}(a) similar as in previous experiments \cite {Kell2004, Tut2004,Wie2004}. We use a dc input current $I$ which is generated by applying a voltage to a large resistor (100 M$\Omega$, not shown) in series to the bilayer system. The current passes first through the top layer, is redirected via a loop resistor (10 k$\Omega$) into the bottom layer in the opposite direction, and finally guided via another resistor (10 k$\Omega$) into ground. The total $I_{Total}$ and the loop $I_{Loop}$ currents are calculated from the voltage drops across the 10 k$\Omega$ resistors. The interlayer tunneling current $I_{Tunnel}$ is  the difference $I_{Total}$ - $I_{Loop}$.

In Fig.~\ref{fig:1}(b) we plot $I_{Tunnel}$ and $I_{Loop}$ as function of the total (dc) input current  for a  $d/l_B$=1.37 at $\nu_T$=1. Strikingly, at small total currents, we find that  $I_{Loop}$ is nearly zero. This is  tantamount to almost all current tunneling before reaching the end of the Hall bar.  With increasing total current, a critical current value  $I_C$=1.5~nA is reached, where  $I_{Tunnel}$ suddenly becomes negligible and nearly all current flows through the loop resistor.  At the same current the longitudinal voltage ($V_{xx}$)  measured along one of the layers  jumps from almost zero to a finite value and continues to increase rapidly with increasing $I_{Total}$ (Fig.~\ref{fig:1}(c)).
Surprisingly, this transition does not indicate the collapse of the $\nu_T$ = 1 state. The magneto-transport measurements presented in the next sections will show that the $\nu_T$ = 1 state changes at $I_C$ into a regime where the dissipation in the longitudinal transport increases with current. {However, below $I_C$ the interpretation of the longitudinal resistance $R_{xx}$ in terms of dissipationless transport is difficult as the current may simply tunnel before passing both voltage probes.}

We estimate from the increase of the loop current at $I_C$ that the effective interlayer resistance must be less than 300 $\Omega$ below $I_C$. The values of  both $I_C$  and the interlayer resistance are comparable to the ones reported in Ref.\cite {Tie2008} where tunneling was measured without a large series resistor. This interlayer resistance is several orders of magnitude smaller than at any other magnetic field. It is also much smaller than in the counterflow experiments of Ref.\cite {Kell2004,Tut2004} where the tunneling never exceeded a few percent of the total current. 

It has been reported in  Ref. \cite {Tie2009} that the value of $I_C$  depends on the effective layer separation $d/l_B$ but also on the area and the bare tunneling probability. The effect of a larger $d/l_B$-value of 1.68 is shown in Fig.~\ref{fig:1} (d) and (e).  In this case, the critical current is only 0.5~nA. In contrast to the sharp jump for $d/l_B$=1.37, tunneling changes more gradually with changing current, a behavior which we observe to be typical for weaker coupling or larger $d/l_B$.

\begin{figure}
\includegraphics{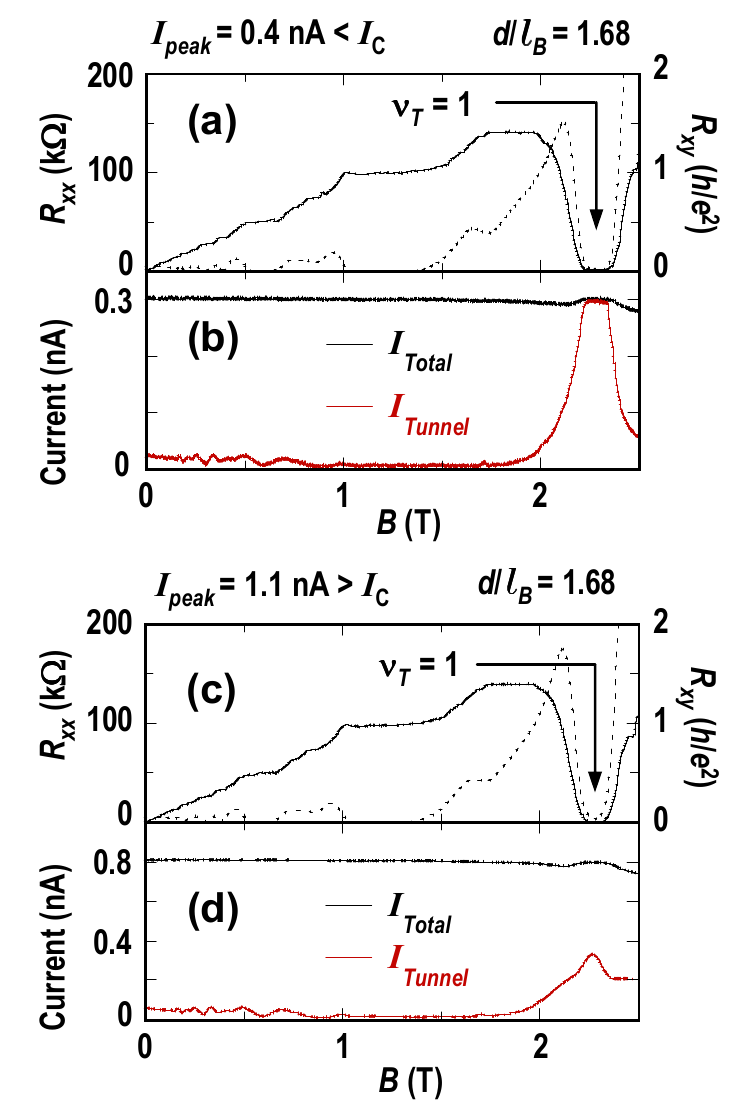}\\
 \textbf{ \caption{\label{fig:2}\textbf{Transport (a and c) and tunnel measurements (b and d) at $d/l_B$=1.68 where $I_C$ = 0.5 nA using ac excitation with a peak current of 0.4 nA (upper two panels) and 1.1 nA (lower two panels), respectively. The minima at  $\nu_{T}=1$ in both the $R_{xx}$ (dashed) and the $R_{xy}$ traces remain pronounced even if the peak current exceeds the critical tunnel current value by a factor of two. Note that rms current values are plotted.}}}
\end{figure}

{Fig.~\ref{fig:2} compares transport (a and c) and tunneling (b and d) measurements as a function of the magnetic field $B$ for $d/l_B$=1.68 with $I_C$=0.5~nA, using an ac excitation rather than dc.} {This allows  standard lock-in technique leading to much less noise and drift than using dc.} {The peak value ($I_{peak}$) of the sinusoidal input current is 0.4~nA $ <I_C$ for Fig.~\ref{fig:2}(a) and (b), while for (c) and (d) $I_{peak}$ is 1.1~nA $> I_C$.}
{With the peak current exceeding $I_C$, tunneling at $\nu_T$=1 is dramatically reduced, but since the sinusoidal input current is less than $I_C$ over part of each period, a finite tunneling current is detected by the lock-in technique. In a pure dc measurement, tunneling would remain small.} {The impact on magneto-transport, however, is only a slight increase of the longitudinal resistance $R_{xx}$. The minimum at $\nu_T$=1 remains. This observation will be discussed more thoroughly with the data of Fig.~\ref{fig:4}.}

\begin{figure}
  \includegraphics{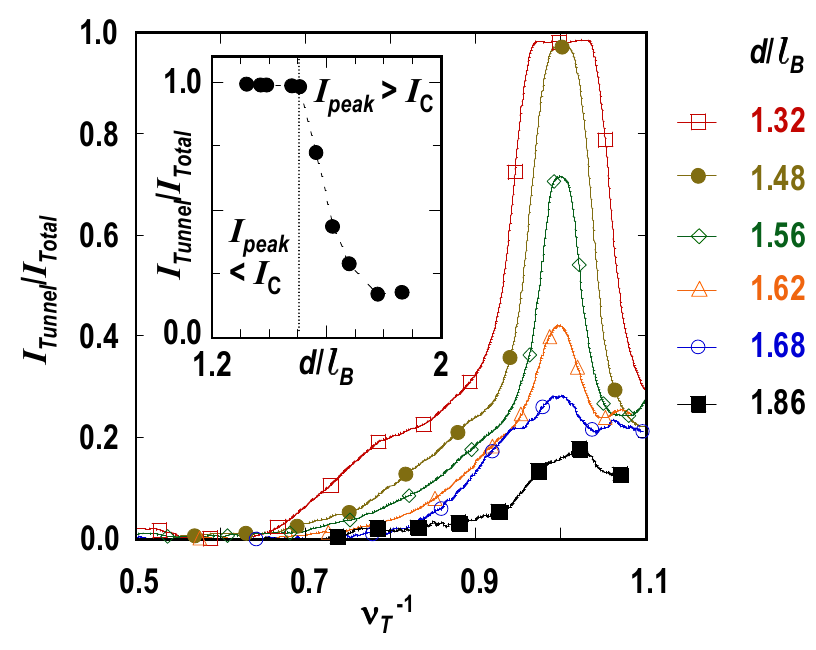}\\
  \textbf{ \caption{\label{fig:3} \textbf{The ratio of $I_{Tunnel}$ to $I_{Total}$ vs $\nu_{T}^{-1}$ with changing $d/l_B$ at a fixed ac current ($I_{peak} \simeq $1.4~nA). Inset: the ratio of $I_{Tunnel}$ to $I_{Total}$ at $\nu_T$=1 vs $d/l_B$. Note that for $d/l_B < 1.50$ the value of $I_{peak}$ becomes less than $I_C$.}}}
\end{figure}

Fig.~\ref{fig:3} illustrates how tunneling varies as a function of the effective layer separation $d/l_B$ by varying $l_B$ for a fixed $I_{peak}$ of 1.4~nA. The traces, obtained by performing the same transport experiment as the one presented in Fig.~\ref{fig:2}, show the ratio $I_{Tunnel}$ to $I_{Total}$, plotted as a function of the inverse total filling factor $\nu_{T}^{-1}$ near $\nu_T$=1. The inset reanalyzes the tunneling at $\nu_T$=1. There is a dramatic reduction of tunneling upon increasing $d/l_B$ to values larger than 1.50 indicating that the values of  $I_{peak}$ and $I_C$ have become identical. Again, the ratio $I_{Tunnel}$ to $I_{Total}$ would have dropped more rapidly if dc instead of ac currents would have been used.

{We will now discuss the dissipation at currents exceeding the critical value in the counterflow set-up. In Fig.~\ref{fig:4} we present several traces of $R_{xx}$ and $R_{xy}$ at  $d/l_B$=1.37 where $I_C$=1.5~nA. These data were taken using a dc current. Each trace corresponds to a sweep of the magnetic field with a different dc-current value. The $R_{xy}$-traces are quite noisy which is typical for our dc-setup. It is striking that upon increasing $I_{dc}$ to above $I_C$, the formerly deep and nearly complete minimum in $R_{xx}$  weakens but does not disappear. Furthermore, $R_{xy}$ remains small and nearly independent of the current (at around 2~k$\Omega$). Both observations are similar to those observed in the earlier counterflow experiments \cite {Kell2004, Tut2004, Wie2004} which had used the ac technique. Our results shows clearly that the transport features which are typical for the $\nu_T$=1 state do not vanish when the critical current is exceeded in the counterflow geometry.
 
There is a peak-like structure in the $R_{xx}$ traces at a  \emph{B} of about 1 T becoming visible at large currents and which is possbly related to forming the $\nu_T$=1 state. It could be due to an enhanced percolation of puddles as suggested in \cite{Stern2002}. Further study is required for its interpretation, however.

\begin{figure}
  \includegraphics{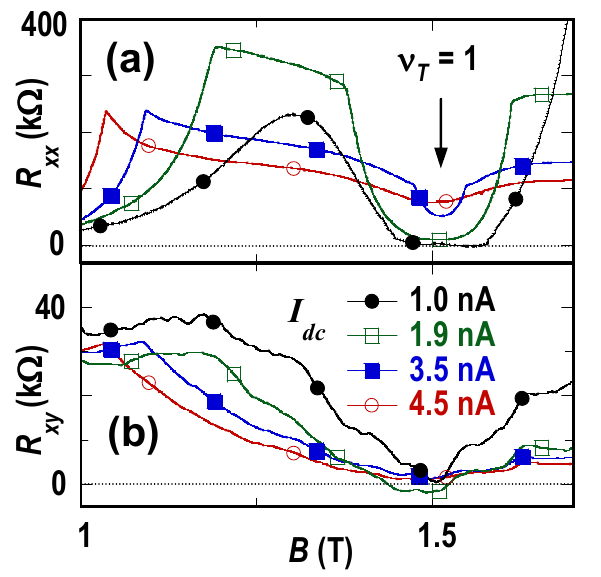}\\
  \textbf{ \caption{\label{fig:4} { \textbf{ $B$ dependence of $R_{xy}$ and $R_{xx}$ at $d/l_B$=1.37 where $I_C$=1.5 nA with dc-currents of 1.0, 1.9, 3.5  and 4.5 nA, respectively. Note that for $I_{dc} > I_C$ the $R_{xx}$ minima become weaker but remain clearly visible. }}}}
\end{figure}

Unfortunately it is not possible to determine $R_{xx}$ in the regime $I < I_C$ quantitatively because tunneling could be occurring between the voltage probes. It could even be argued that the current takes a "tunneling-shortcut" to the other layer, leaving the voltage probes in a currentless part of the sample. Thus we cannot state that  the longitudinal dissipation is near zero at $I < I_C$. On the other hand, it has been recently reported that tunneling occurs over the total sample area \cite{Fin2008,Tie2009} meaning that we should have observed finite $R_{xx}$ values in the  $I < I_C$ regime if there was a sizeable dissipation in the strong tunneling regime. 

These findings show that exceeding the critical tunnel current does not cause the breakdown of the  $\nu_T$=1 quantum Hall state in the counterflow set-up.  Previous counterflow experiments required a connecting loop to observe the nearly vanishing of $R_{xx}$ and $R_{xy}$. The small tunneling observed in those experiments was most likely caused by the fact that the critical currents were smaller than the respective measuring currents. 

While exceeding the critical current for tunneling does not destroy the condensed exciton state, it seems to mark the onset of dissipative processes which increase rapidly with increasing current. 
Recent theoretical suggestions attribute the dissipation in these exciton condensates to the unbinding and the current-induced motion of vortices connected to charged quasiparticles (merons)  \cite {Fer2005, Hus2005,Roo2008,Eas2009}. It is, however, not clear why this process should become much more effective at $I > I_C$.    In another recent  theory it has been proposed that finite voltages between the layers produce magnetic vortices parallel to the layers which are forced to move due to the same voltage difference \cite {Fil2009} . This scenario would lead to a sudden onset of dissipation and is therefore quite attractive to describe our findings.

In summary we have demonstrated that a counterflow experiment in an electron bilayer system at $\nu_T$=1 shows  the typical transport signatures of the expected excitonic superfluidity {up to currents several times larger than the critical current for tunneling}. Below $I_C$, coherent tunneling is active and is causing an "electrical short"  making the external loop wire connecting both layers in series redundant.  On the other hand, above $I_C$, the large tunneling resistance is restored and dissipation in a counterflow experiment begins to increases rapidly with current. 

We would like to acknowledge discussions and correspondence with 
 P. R. Eastham,
 D. V. Fil,
 M. Gilbert,
 A. H. MacDonald, 
 C. de Morais Smith and 
 B. Rosenow.
 The samples were produced in collaboration with H.-P. Tranitz  and J. G. S. Lok. 
This project was supported by the BMBF (German Ministry of Education and Research) Grants No. 01BM456 and 01BM0900.\newline


\end{document}